\documentclass[%
 reprint,
 amsmath,amssymb,
 aps,nofootinbib,   
]{revtex4-1}
\usepackage{bm}
\PassOptionsToPackage{linktocpage}{hyperref}
\usepackage[hyperindex,breaklinks]{hyperref}
\usepackage{enumitem}
\usepackage{slashed}

\renewcommand{\theta}{\vartheta}

\usepackage{array}
\usepackage{mathtools}

\usepackage{etoolbox}

\begin{document}

\title{Cosmological Relaxation of Higgs Mass Before and After 
LHC \\
 and Naturalness }

\author{Gia Dvali} 
\affiliation{%
Arnold Sommerfeld Center, Ludwig-Maximilians-Universit\"at, Theresienstra{\ss}e 37, 80333 M\"unchen, Germany, 
}%
 \affiliation{%
Max-Planck-Institut f\"ur Physik, F\"ohringer Ring 6, 80805 M\"unchen, Germany
}%
 \affiliation{%
Center for Cosmology and Particle Physics, Department of Physics, New York University, 726 Broadway, New York, NY 10003, USA
}%

\date{\today}

\begin{abstract} 
 In post LHC era the old idea of cosmological vacuum relaxation of the Higgs 
 mass that does not require any new physics in the vicinity of 
 LHC energies acquires a new meaning. 
 I discuss how this concept of naturanless differs from the 
 standard one by 't Hooft. 
 Here the observed value of the Higgs mass 
 corresponds to a vacuum of infinite degeneracy and infinite entropy.
 Therefore, it represents and attractor point of cosmic inflationary evolution.  
  This information is unavailable for a low energy observer living in 
 one of such vacua.  By not seeing any stabilizing physics at LHC 
 such an observer is puzzled and creates an artificial problem of  naturalness which in reality does not exist.  We 
 explain why this solution is fully compatible with the concept 
 of Wilsonian decoupling.   
 \end{abstract}

\maketitle
\section{What is naturalness?}

 The main motivation (if not the only one) for seeing the new physics 
 at LHC was and is the {\it Hierarchy Problem}.  This is a tremendous
 puzzle  that spans scales over $34$ orders of magnitude. 
   Gravity plays the key role in 
 defining this problem.
 The reason, as explained in details 
in \cite{Gia}, is that gravity introduces the Planck mass $M_P$ that 
separates the worlds of elementary particles and black holes. This makes the hierarchy problem real. \\

 In a world without gravity, one could take an attitude 
that in a renormalizable theory the quadratic sensitivity of the Higgs mass
towards the cutoff scale $M$  is 
an artefact of perturbation theory. At the end of the day,  what matters are the measurable physical parameters. They are what they are and we should not worry about explaining their smallness.  So, if a renormalizable theory contains a single input scale, it is meaningless to ask why it takes a particular value.   \\

  In the presence of gravity this argument fails because of the following reason \cite{Gia}. The point is that no elementary particle can exist 
  with its mass $m$ largely exceeding $M_P$. Any such object becomes an extended {\it classical}  object with its localization radius 
  bounded from below by its gravitational radius $R_g = m/M_P^2$. 
 For $m \gg M_P$  this radius exceeds the 
 Compton wavelength of the ``particle" in question, $R_C \gg 1/m$. This signals that the object we are dealing with is no 
  longer microscopic and quantum but rather is macroscopic and (semi)classical. 
   \\

Due to above, 
some hierarchies are more puzzling than the others. 
 For example, we are not bothered by the question why is the Higgs much lighter than the Sun.  This is because we intuitively understand that
in the opposite case, the  Higgs boson would not represent an elementary particle but
instead a classical  black hole with solar mass! 
 Within the framework of the Standard Model plus Einstein gravity 
 no such argument exists for explaining why the Higgs boson  cannot have 
 a nearly Planckian mass.  
Thus, we, as low energy observers, are puzzled by the fact that the Higgs boson  is much lighter than the Planck scale. 
 We think this is {\it unnatural}. \\
 
    {\it So, what is naturalness? } \\
    
   For answering this question let us first understand whether and how can we distinguish a naturalness problem 
from a consistency one. For example, take the hypercharges of quarks and leptons in the Standard Model. It is impossible to judge their assignment     
 by naturalness. Instead, the hypercharges of fermions  are fixed by the anomaly cancellation which is a {\it consistency}  requirement.  So, how do we know that the Higgs mass is not  fixed by some underlying consistency
 requirement? Well,  we do not. All we can say is that this does not 
 appear to be the case within the Standard Model understood as an effective low energy theory. \\
 
 If we think deeper, we may realize that in our judgement we could be dealing with the power of precedent. 
Namely, the reason why usually we do not 
 classify  the smallness of the Higgs mass as a consistency issue, is that 
 the solutions do exist 
 that can explain it  {\it naturally}.  That is, ``subconsciously",
 we use the existing solution as a reference point.   
Thus,  in some sense, 
 when we attribute a certain puzzle to a naturalness category, we reason  
backwards, abstracting from some ``sample" solution. 
The representative examples of such sample solutions are provided by the low energy supersymmetry, by Technicolor \cite{TC}  and  by large extra dimensions \cite{ADD,RS}.  
\\

On the other hand, when the search for a natural solution to a given puzzle fails after many attempts, 
we can start thinking that the answer may hide in consistency 
rather than in naturalness. The example of the latter situation is provided 
by the Cosmological Constant Problem \cite{Weinberg}.  Literally, all the attempts of solving 
this problem in terms of naturalness failed miserably. 
This fact prepares a serious ground for moving this problem 
from the naturalness category into the consistency one \cite{GiaCesar}. \\

Another thing we can say is that we always recognize 
naturalness whenever we encounter it. 
Again, this encounter usually comes {\it post factum} after we discover  an 
explanation to a phenomenon that we adopt as  ``natural". 
This is especially true when the underlying mechanism is elegant. 
For example, in the light of asymptotic freedom, we accept that having proton much lighter than the Planck 
mass is natural because the QCD gauge coupling 
runs logarithmically with energy. Therefore,
it takes a descend to very low energies for the QCD to become strong and form bound-states such as proton.

\section{High Entropy Vacuum is Natural}  
  
 There exist cases in which the concept of naturalness can be defined very precisely. One is the ``old" naturalness in the sense of  't Hooft \cite{tHooft}.  According to this definition, a value $\Phi_*$ of a 
 physical quantity $\Phi$ is natural if this value corresponds to an enhanced 
 symmetry point:
\begin{equation} \label{thooft} 
  \Phi \rightarrow \Phi_*  \, ~~ = ~Enhanced~symmetry  \, .
 \end{equation}
 This concept is  extremely well-known and sometimes is referred to as 
 {\it technical naturalness}. This is because radiative corrections are controlled by symmetries, at least in the absence of anomalies.  \\
 
  The application of the above  notion of naturalness to the Hierarchy Problem leads us to postulate a new symmetry that protects
 the Higgs mass against quantum corrections. A well known example
 of this sort is provided by the low energy supersymmetry.  In this context, the  naturalness {\it a la}  't Hooft applied to the solution of the Hierarchy problem motivates a search for new 
 physics at LHC in form of the super-partners of the Standard Model particles.  \\

  We shall focus on a very different concept of naturalness introduced
 and applied to the Hierarchy Problem in pre-LHC era (Dvali, Vilenkin
 '03, Dvali '04,  \cite{attractor}).
  This naturalness can be defined as {\it enhanced entropy of the vacuum}.   
  That is, a value $\Phi_*$ of a 
  physical quantity $\Phi$ is natural if the theory contains 
  a sharply enhanced number $n_V$ of vacua with $\Phi = \Phi_*$. 
  In such a case we can define the {\it vacuum entropy}  as 
  \begin{equation}\label{VE}  
  S_V  \equiv 
  \ln(n_V)\, .  
  \end{equation}
  Thus, the natural value $\Phi_*$ corresponds 
  to the enhanced vacuum entropy point.  
 Putting it shortly:
 \begin{equation} \label{thooft} 
  \Phi \rightarrow \Phi_*  \, ~~ = ~{\rm Enhanced~Entropy} \, . 
 \end{equation}
   The essence of the naturalness defined in this way is intuitively clear. 
   After all, the systems that we observe in everyday life either already exist in their maximal entropy states or evolve towards them. 
   Therefore, if we succeed in constructing a theory in which the 
 vacuum with a small weak-scale-to-Planck-scale ratio has a maximal entropy, it would be natural to find ourselves in such a vacuum sector.  \\  
  
  However, the two things must be distinguished. 
  First is that the theory must posses 
  such an enhanced entropy vacuum. Secondly, there must exist a cosmological mechanism 
  that allows us to populate it.   In the original construction
 (Dvali, Vilenkin, '03;  Dvali, '04  in  \cite{attractor})  the latter was achieved via the cosmological relaxation of the Higgs mass during the eternal inflation. 
  In the process of this relaxation the Higgs mass acts as a feed-back control parameter dictating to which vacuum 
  the cosmic evolution must converge.  That is, the Higgs mass is dynamically promoted  into a 
self-controlling stochastic variable. \\

 The  novel and a potentially puzzling feature of the above scenario is that the hierarchy 
 is made natural thanks to new physics that can be arbitrarily heavier than the weak interaction scale.  A legitimate question is how is this story reconciled 
 with the effective field theoretic notion of Wilsonian decoupling.  
  Before making this clear, we shall briefly summarize the essense of the solution.

   \section{Back-Control Mechanism}
    
   Since all the subsequent scenarios in \cite{attractor}  are based on the same 
   general idea and mechanism, we shall illustrate the point by closely following  the original construction of (Dvali, Vilenkin, '02;  Dvali, '05).  
  Also, the advantage of this version is that it offers a maximal hierarchy. 
  That is, the scale of new physics $M$ can naturally be pushed all the way to $M_P$.  \\
    
    The model incorporates the following two aspects: 
    \begin{itemize}
  \item 1) Statics:  The vacuum with the hierarchically-small Higgs VEV represents an {\it infinite entropy}  point in the space of vacua;  

  \item  2) Dynamics:  By cosmological evolution due to eternal inflation 
  the Universe is driven towards this infinite entropy point.  

\end{itemize}
        
   In this way, the desired vacuum with a strong hierarchy between the 
   weak and the Planck scales becomes a dynamical attractor point of  the cosmic evolution. 
    The mechanism is realized using the following three ingredients: 
 
  \begin{itemize}
     
    \item 1) A scalar $\Phi$ which  represents  a prototype of SM Higgs
    (for simplicity we take it real); 
   \item  2) A massless $3$-form field  $C_{\alpha\beta\gamma}$ 
    with corresponding $4$-form  field strength 
    $F_{\alpha\beta\gamma\delta} \equiv 
    \partial_{[\alpha} C_{\beta\gamma\delta]}$  
   and its Hodge-dual scalar, $F \equiv \epsilon^{\alpha\beta\gamma\delta}F_{\alpha\beta\gamma\delta}$;

 \item   
    3)  A $2$-brane that is sourcing $F$. The strength of the source, i.e., the $2$-brane charge $Q(\Phi)$,   is controlled by $\Phi$.
 Alternatively, the $2$-brane can be replaced by an axion-like field
    $a$.  Both ingredients in \cite{attractor} were taken from the earlier work \cite{DV}.  
 
 \end{itemize}

    The three entities control each other in the following way. 
   The VEV of the Higgs field $\Phi$ in each vacuum depends on $F$.   
 Since $F$ is sourced by a $2$-brane, we can move to a different vacuum provided we cross the $2$-brane that separates the two vacua. 
 During this move $F$ experiences a jump given by, 
 \begin{equation}\label{jump}
 \Delta F = Q(\Phi) \,.  
 \end{equation} 
 This  jump triggers the respective change of the
Higgs VEV, $\Delta \Phi$.  
 In this way, the  $\Phi$-dependence of the brane charge $Q(\Phi)$ back-controls the subsequent variation of the Higgs VEV. That is, if we decide to move 
 to a third vacuum, the new step now will be given by 
 $\Delta F = Q(\Phi + \Delta \Phi)$.    
Thus, the magnitude of the jump $\Delta F$ decreases in the direction in which $|Q(\Phi)|$ decreases.  It is clear that if for some value of the Higgs VEV $\Phi_*$
 we have $Q(\Phi_*) = 0$, then  
  reaching the point $\Phi_*$ requires an infinite number of jumps. 
  Thus, the vacuum $\Phi_*$ is separated from any 
 infinitesimally close value 
  $\Phi \neq \Phi_*$ by an {\it infinite}  number of vacuum states.  
  That is, the vacuum $\Phi_*$ exhibits an infinite degeneracy and an infinite entropy. \\
  
  In order to make this more explicit consider the following Lagrangian \cite{attractor},        
   \begin{eqnarray}   \label{L1} 
      L &=&   (\partial_{\mu} \Phi)^2\,  +\,  F^2 + \,   \\ \nonumber
    && + \, \left ({F^2 \over M^2}  - m^2\right ) \Phi^2 \,  - \, {\lambda \over 2}\Phi^4  -\\ \nonumber
    &&\, - \,  J^{\mu\nu\alpha}_T C_{\mu\nu\alpha} \,, 
    \end{eqnarray}  
 where $M$ is a cutoff and $m$ parameterizes all possible contributions 
 to the Higgs mass in the absence of $F$.  Of course, since 
 we wish to solve the Hierarchy Problem we shall assume 
 that $m$ is naturally of order the cutoff $M$. We then show 
 that in the attractor vacuum the effective Higgs mass will be almost cancelled 
 by the contribution coming from $F$. 
 In (\ref{L1})  a simple coupling among Higgs and $F$ has been chosen. The mechanism is universal 
 and works for a generic form of $F$-dependence.   
  The quantity $J^{\mu\nu\alpha}_T$ stands for  the transverse part of the following
 $2$-brane-current, 
  \begin{equation} \label{current}
  J^{\mu\nu\alpha}\equiv Q(\phi)\int d^3\xi  \delta(x-Y(\xi))
  {\partial Y^{\mu} \over \partial \xi^a} {\partial Y^{\nu} \over \partial \xi^b}{\partial Y^{\alpha} \over \partial \xi^c} \epsilon^{abc} \, ,
 \end{equation} 
 where $\xi^{a} \,~(a =0,1,2)$ are the brane world-volume coordinates  
 and $Y^{\mu} \, ~(\mu=0,1,2,3)$ are the embedding coordinates 
 that parameterize the $2$-brane history from the point of view of 
 $3+1$-dimensional space-time with coordinates $x^{\mu}$.
 In above expressions the obvious combinatoric factors are assumed but not displayed for compactness.\\
  
Due to sourcing of $F$, the brane separates domains with different values of the vacuum energy. This fact has been used in the past 
for a neutralization of the cosmological constant via brane nucleation (Brown, Teitelboim, '87 \cite{BT}).
 The idea of \cite{attractor}, is to use the analogous process for scanning the Higgs mass due to the coupling 
 between $F$ and $\Phi$.  The difference from cosmological constant is that the 
 scanning happens in a {\it self-controlled}  manner.   
  \\ 
 
  The key point is that the effective charge of the brane $Q(\Phi)$ 
  depends on $\Phi$ in such a way that 
  $Q(\Phi_*)=0$. \footnote{Since $Q$ is not a constant, 
  $3$-form gauge invariance demands that $C$ is sourced by 
  the transverse part of the current (\ref{current}). This is achieved by 
  integrating out a St\"uckelberg fields as shown in  (Dvali, '04 \cite{attractor}).} 
  
  For example, 
  \begin{equation} \label{Q}
  Q(\Phi) = \pm {1 \over M^{NK -2}} (\Phi^N - \Phi_*^N)^K  \, , 
  \end{equation} 
where $N$ and $K$ are some positive integers and $M$ is a cutoff scale 
that can be arbitrarily high and all the way up to $M_P$.  
The quantity $\Phi_*$ is of order  $100$GeV.  Note, 
this value of $\Phi_*$ emerges as an interplay 
between the cutoff  $M$ (Planck mass) and QCD condensate of quarks
(see, Dvali, '04 in \cite{attractor}). For the details of the construction we refer the reader to the original paper. 
For the present discussion we take this value of $\Phi_*$ for granted.    
Note, this quantity is stable with respect to radiative corrections. \\

  The vacuum with corresponding 
  Higgs VEV $\Phi_*$ is dynamically selected by the attractor mechanism as follows.    
 By minimizing the Higgs potential in (\ref{L1}),  the VEV of the 
 Higgs as a function of $F$ is given by, 
 \begin{equation} \label{VEV}
   \Phi^2 = {1 \over \lambda} \left ( {F^2 \over M^2} - m^2 \right )\, .
     \end{equation}   
  We assume for definiteness that all parameters (and of course the r.h.s. of the equation)  are positive. 
 Once we move into a neighbouring vacuum across the $2$-brane, 
 $F$  changes by (\ref{jump}).
  The corresponding change of the Higgs VEV from  
  (\ref{VEV}) is, 
 \begin{equation} \label{DVEV}
   \Phi \Delta \Phi = {1 \over \lambda} {F \over M^2}  Q(\Phi) \, .
    \end{equation}
    Since we are interested in the vacuum structure near the attractor point, 
  terms up to linear in $Q(\Phi)$ have been kept.   
 Also, for definiteness we assume that we are approaching the attractor from above, meaning that 
 $\Phi$ and $F$ are decreasing and $Q(\Phi)$ is negtive. 
 From here, we can easily estimate the density of vacua in some small 
 neighbourhood of the attractor point $\Phi_*$.  
 Since  near the attractor $Q(\Phi)$ is almost zero, the VEVs  
 $\Phi$ and $F$ are nearly frozen and their variations from vacuum to vacuum  are infinitesimal.  
   So the number of  vacua within a finite  
   interval  between  some $\Phi_0$
and  $\Phi_* + \epsilon$, where $|\Phi_0 - \Phi_*|  \gg \epsilon$, 
can be estimated as, 
     \begin{equation}
  n_{V} \simeq  \lambda {M^{2} \over F_*}  \int^{\Phi_* + \epsilon}_{\Phi_0}
 {\Phi d \Phi \over Q(\Phi)} \,,
   \end{equation}
   where 
 \begin{equation} \label{Fstar} 
  F_* = M \sqrt{\lambda \Phi_*^2 + m^2} \,, 
  \end{equation} 
    is the value of $F$ for $\Phi = \Phi_*$. 
   Since near the attractor point $F$ remains uniformly large 
   ($\sim M$ ) and constant,  we moved it out of the integral. \\ 

  For any finite interval $|\Phi_0 - \Phi_*|$  this expression diverges for 
   $\epsilon \rightarrow 0$. 
  For example, taking  $Q(\Phi)$ as in  (\ref{Q}), 
    we obtain 
    the following form of the divergence of the number of vacua: 
    \begin{equation} \label{divergence} 
  n_V|_{\epsilon \rightarrow 0}  =   \begin{cases}
    c {1 \over (K-1) \epsilon^{K -1}} & \text{for}~ K> 1, \\
    c\ln 
      \left ({1 \over  \epsilon } \right )  & \text{for}~K=1, 
\end{cases}
  \end{equation} 
  where the constant coefficient is given by $c=\left ( {\lambda  M^{NK} \over F_* \Phi_*^{NK -K -1}N^K}  \right )$.     
 Note that in the limit $\epsilon \rightarrow 0$ the dependence 
 on $\Phi_0$ drops out due to the divergent number of vacua 
 around the attractor point 
 $\Phi_* $. \\

  Thus, the attractor mechanism generates infinite 
  entropy vacuum with the Higgs VEV given by $\Phi_*$,  
  \begin{equation} \label{entropy}
  \lim_{\Phi \rightarrow \Phi_*}   S_V =  \infty \,.
    \end{equation}   
  
 Couple of comments are in order. First, the reader should not be 
 frustrated by the notion of {\it infinite entropy}.  The objects 
 of finite radius cannot have an infinite entropy \cite{Bek} but vacua 
 can. \\
 
  Secondly, it is obvious that the attractor vacuum is stable
 at zero temperature.  Namely, the existence of infinite number of closely-spaced vacua in the field space is not increasing 
the decay probability.  In fact, the instability rate is negligible since the bubble-nucleation only gives transitions between the neighbouring vacua. 
The difference between the energy densities of the neighbouring vacua can be estimated as 
$\Delta {\mathcal E} \sim  Q(\Phi)F$.  Then, for the brane tension 
given by the cutoff scale $\sim M^3$ the Euclidean action of the
bounce that describes tunnelling is  
$S_E \sim M^6/Q(\Phi)^3$. This makes the transition around the attractor point  $\Phi_*$ essentially impossible at zero temperature. \\

Of course, during inflation the story is very different. 
 As shown in (Dvali, Vilenkin '03 \cite{attractor}),    
 the attractor vacuum is actualized by the mechanism of eternal inflation
 (Vilenkin '83 \cite{eternalV}, Linde '86 \cite{eternalL}). 
 Indeed, during inflation the transition among different vacua take place 
 with non-zero rate regardless of the brane tension separating these vacua.
 If the brane tension is of order $M^3 \sim M_P^3$, the 
 production rate of the bubbles is $\sim {\rm e}^{-\pi {M_P^2 \over H^2}}$ 
 where $H$  is the inflationary Hubble parameter (Basu et al, '91 \cite{rate}). 
 Correspondingly, since inflation lasts eternally, the Universe has 
infinite time for exploring the entire vacuum landscape.  In this situation 
most of the Universe is expected to be driven towards 
the attractor vacuum due to its infinite entropy.  \\ 

Finally, as a technical remark.  Having an explicit $3$-form field is not essential. As was shown already in (Dvali, Vilenkin '03, Dvali '04 
\cite{attractor}), we can  
resolve the brane as a domain wall of an axion-type 
field $a$ with a periodic potential $V(a)$, say with periodicity 
$2\pi$. 
In this case the brane current becomes, 
 \begin{equation} \label{resolve}
  J^{\mu\nu\alpha} \rightarrow  Q(\Phi)\epsilon^{\mu\nu\alpha\beta}\partial_{\beta}
  a \, .
 \end{equation}  
 Next, integrating out the $3$-form results into a coupled Higgs-axion system. The  stationary points of it (at the leading order)   
 are determined by the following equations 
(Dvali '04, \cite{attractor}) 
   \begin{eqnarray}   \label{EFF} 
    &&  \left (a^2 {Q(\Phi)^2 \over M^2} - m^2 \right ) = \lambda \Phi^2  \\ \nonumber
    &&  d_aV(a) + Q(\Phi)^2\, a =0 \,,  
    \end{eqnarray} 
  where for clarity the numerical factors have been rescaled properly and integration constant has 
been set to zero.   
The last equation gives the effective potential for the axion field. 
The remarkable thing about this potential is that it  becomes exactly periodic  asymptotically for
 \begin{eqnarray}   \label{LIM} 
     && \Phi \rightarrow \Phi_*   \\ \nonumber
    && a \rightarrow {M \over Q(\Phi_*)} \sqrt{ \lambda \Phi_*^2 + m^2} = \infty \,.  
    \end{eqnarray} 
 This is a manifestation of infinite degeneracy 
of vacua with $\Phi  \rightarrow \Phi_*$. This version of relaxation 
was further adopted in several subsequent papers in \cite{attractor} (see, e.g.,
Graham et al '15, Espinosa et al, '15, Chiara et al, '15).\footnote{Notice \cite{DV}, in the absence of back reaction on $\Phi$ the axion potential obtained 
after integrating out the $3$-form would have a form of ``washboard"  of the type that was used by Abbot \cite{abbot} for the relaxation of the cosmological constant. 
Here, the level-splitting 
that is proportional to $Q(\Phi)$ can be made naturally small by
symmetries of $\Phi$ \cite{DV}. 
However, with such a potential the Hierarchy problem would not be solved since the Higgs mass would not act as a self-control parameter. 
Instead, it had to be determined by some other physics.  
This problem is solved in  \cite{attractor} where  the crucial novelty is that  not only $Q(\Phi)$ depends on the VEV of the Higgs field but  at the same time the latter is back-controlled  due to the coupling 
between $F$ and $\Phi$.}

 \section{Decoupling} 
 
 One of the most fundamental concepts of effective field theory  is
Wilsonian decoupling:   High energy physics must decouple from the low energy one.  Misinterpretation of this concept may create a 
certain prejudice against the 
 possibility of dynamically solving the Hierarchy Problem 
 without employment of new physics around TeV energies.  \\

  Indeed, at the first glance it is hard to imagine a dynamical adjustment mechanism  that eliminates the cutoff-sensitivity of the Higgs mass and relaxes it to a value of order $100$ GeV, if the new degrees of freedom are as heavy as the cutoff.  So, how does the cosmic attractor mechanism defies this intuition? \\
  
  The answer is very simple: { \it Eternity}. \\
  
  The point is that even a very high energy physics can have extremely important low energy effects over sufficiently long time scales, especially 
in cosmological context.   
  
   In order to understand this, it is enough to recall the phenomenon of proton decay in grand unification. 
  Indeed,  in grand unified theories the exchanges of heavy particles mediate very low energy processes such as proton decay.   Despite the fact that the corresponding effective high-dimensional operators 
are extremely suppressed,  they have vital effects over sufficiently 
long time-scales and/or for high occupation number of particles. 
This is the reason why the experimental searches for proton decay
require large tanks of water:  We need to have a large number of samples. 
 However, to gather many samples we need to employ some low energy interactions
 with long-range correlations. In other words, a long waiting time and/or  a high occupation number  
 can overpower the  decoupling of heavy physics. \\
  
  The vacuum relaxation mechanism for the Higgs mass \cite{attractor}  works in a similar way.  Indeed, the physics that leads to the 
existence of the high entropy Higgs vacuum involves heavy degrees of freedom. 
  These are branes that have a tension given by the cutoff scale 
  $M$, or equivalently,  an axion type field of a high mass. 
 However, for solving the Hierarchy problem we need to actualize 
 enormous number of vacua. This is like gathering the large tanks of water.   
  For achieving this goal, a physics with long range correlations is required. 
  There indeed exists such a degree of freedom that 
  actualizes all possible vacuum states and in this way brings the effect of heavy physics down to low energies.  This degree of freedom is the
  {\it massless graviton}.  
  \\
  
   Indeed, a short-time effect of the heavy branes at low energies is absolutely negligible. 
  For example, a zero temperature observer 
 has no chance of reaching the attractor vacuum by the brane nucleation.
  However, gravity, through the eternally lasting inflation, changes the story dramatically. 
 Although the hot big bang epoch of our observable Hubble patch is still relatively young,  it was preceded by the inflationary era. 
 According to the concept of {\it eternal inflation} 
 (Vilenkin '83, \cite{eternalV}, Linde '86 \cite{eternalL}). 
  The latter era lasts
 permanently by the mechanism of self-reproduction. In such a scenario 
 the Universe has an unlimited time to explore all possible vacuum 
 domains.  
  During eternal inflation the vacuum landscape is scanned due to permanent nucleation of the vacuum bubbles. Since the inflation is lasting eternally, the  Universe is inevitably driven to the attractor point due to its infinite entropy. \\
  
  In other words, it is the existence of the massless graviton that reconciles 
  the attractor solution to the Hierarchy Problem with the notion of 
  the decoupling.  That is, gravity, thanks to inflation, is capable of driving the Universe towards  the highest entropy vacuum state. 
  In this sense, the  actualization  of the attractor vacuum  by eternal inflation can be 
 spiritually compared with the formation of a black hole during a gravitational 
 collapse or in any other very high energy process.  There too 
 the system is driven towards a maximal entropy state that is represented 
 by a black hole. \\
 
 The latter connection is deeper than what one may naively think.
 Indeed, in order to be able to form an arbitrarily large black hole it is absolutely crucial that gravity is both gapless and universally coupled.
 To put it differently, the equivalence principle makes Einstein gravity an universal actualizer. 
 That is, in order to actualize 
the black hole states of arbitrarily high entropy, the  graviton must be massless.  Like-wise, the massless graviton is crucial for maintaining the eternally inflating Universe. The latter is necessary for actualizing the infinite entropy  Higgs vacuum that solves the Hierarchy Problem.     
 Thus, in both cases the gaplessness and universality of gravity plays the crucial role in actualizing the states (either black holes or vacua) of maximal entropy.

 \section{Discussions}
 
 The cosmological vacuum relaxation mechanism \cite{attractor} offers an approach to the Hierarchy problem 
 that defies the standard view about naturalness.
  It shows that physics that solves the Hierarchy Problem can be arbitrarily heavy 
 and correspondingly arbitrarily-weakly interacting with the Standard Model 
  particles provided the theory contains gravity.  
    Reconciliation of this fact with the intuition 
  coming from the concept of decoupling lies in arbitrarily long time-scales offered by the cosmology of eternal inflation. The role of the 
degree of freedom that connects deep UV and IR physics  is assumed 
by graviton.     
   \\ 
  
   In this scenario, the observed  hierarchically small value of the weak interaction scale  is due to the fact that 
  our vacuum is singular. Namely, it represents an infinite entropy point among all possible vacua. 
  That is, there exist infinite number of degenerate vacua with 
the values of the Higgs mass infinitely close to the observed value.   
   Gravity via the phenomenon of  {\it eternal inflation}  acts as an universal actualizer.   
 As a result  of the eternal inflation, the 
 Universe is driven to the critical value of the Higgs VEV with an unit probability.   
  Correspondingly, the low energy observer 
  cannot observe experimentally any stabilizing new physics  below the scale $M$.  Not seeing so far  any new physics at LHC may be a signal for taking the new concept of naturalness seriously. 
  It may be an indication that a hierarchy between the weak scale and a cutoff originates from the high entropy of our vacuum 
 and our long history of cosmological relaxation. \\
 
 Again we wish to reiterate that the presented concept of naturalness \cite{attractor} relies on 
 gravity as {\it universal actualizer}. Therefore, if we are willing to accept the concept of the eternal inflation
 there is no reason why the concept of naturalness of an infinite entropy attractor vacuum should not be taken seriously and explored further. \\    
 
 We would like to mention few words about predictions. 
 Needless to say, this is an important question for any solution of the Hierarchy Problem. 
  Here we wish to stress two points. First, is about  separating the conceptual issues   
from the model-dependent  ones.  The main thing about the presented ideas is that they encode  
a conceptually different view of naturalness.  The ingredients used in the original work  in \cite{attractor} 
such as  $3$-forms, branes and axions should be viewed as tools for establishing the point that is far more general. 
This is the idea that naturalness of low energy observables may be governed by UV-physics
 through the structure of the vacuum and its cosmological actualization. 
 The main goal of the present paper was to reiterate this message  in the new  light of 
LHC not seeing (so far) any stabilizing physics for the Higgs mass: \\

 {\it Not seeing new physics at LHC does not mean that nature is unnatural}. \\
 
   Rather the naturalness can be realized in the way that is far more profound and hidden from the low energy observer. \\
   
As for quantitative predictions, they are always made within the specific models and emerge as a result  of 
correlations  among distinct phenomena. For example, within a specific model
that makes number of explicit assumptions, such as R-parity and structure of hidden sector,   the scale of supersymmetry breaking may be 
correlated with the mass of a dark matter particle.  
 In this light, the presented approach to naturalness could be equally promising. For example,  one may try to  
correlate the solution to the Hierarchy Problem with that of the strong-CP puzzle. 
In this respect we note that in  \cite{CP} the concept of enhanced-entropy naturalness was applied to the $\theta$-parameter of QCD. 
There an attempt was described in which the vacuum with  the vanishing QCD $\theta$-angle 
has infinite entropy.  This is possible thanks to the fact that QCD $\theta$-vacua can be reformulated 
in the language of a $3$-form gauge theory using the Chern-Simons $3$-form \cite{3F}. We refer the reader 
to the latter paper  for the detailed discussion of this picture.   In this language, 
the different super-selection sectors correspond to vacua with different values of the 
``electric" $4$-form field strength $F$. Obviously,  the  CP-conserving vacuum ($\theta =0$)
  is the one with $F=0$.  
Then, the conventional Peccei-Quinn scenario \cite{PQ} can be 
understood rather elegantly as the effect of Higgsing
the $3$-form gauge field:  The $3$-form eats up a Weinberg-Wilczek axion \cite{axion}  and becomes massive. 
Consequently, the electric field $F$ is screened.  
This results 
into $F=0$ as an unique vacuum.  The idea of \cite{CP}  is  not to make $F$ massive. 
Instead, the strategy is to use the attractor mechanism analogous to \cite{attractor} for promoting  
the CP-conserving vacuum $F=0$ into an infinite entropy point. 
Again, via eternal inflation this vacuum is expected to be actualized with a high probability. 
In such a case, similarly to the Higgs mass in \cite{attractor},  the vacuum   $\theta=0$ is 
natural although a low energy observer may not appreciate this.

Now, combining the vacuum attractor relaxation mechanisms for the Higgs mass and for  $\theta$
can produce the set of correlations and correspondingly new predictions such as the presence of excitations of brane modes and/or axion like fields at low energies.
Their discussion is beyond the focus of the present paper.   \\

 Goran Senjanovic \cite{Goran} was telling me over the years that we should stop worrying  about the Hierarchy Problem as a priority and should keep exploring new physics without any reference to naturalness. 
 And, although the issue is fundamental, it does not harm the Standard Model which is a self-consistent 
 high precision theory.  
The idea of \cite{attractor} discussed in this note provides a full support to this view as  
it liberates a low energy model builder from the burden of
conforming to norms of naturalness. 
Of course, this liberation has nothing to do with a trivial dismissal of the issue.  
   Indeed, the Hierarchy Problem 
 is {\it real} and the mechanism of \cite{attractor} is addressing it in 
 a very precise sense of naturalness.  The only caveat is that the naturalness is hidden for the low energy observers like us.
  The solution is visible 
 only for an observer that can either last eternally 
 or be able to perform very high energy experiments, or both. 
 Therefore, in order to falsify  
 this solution experimentally we need to explore the Planck scale physics as well as understand cosmology on eternal time scales.  This is a perfectly achievable and a noble  
 task but outcome does not boil down to seeing new particles around TeV energies. In a sense the ``new" low energy particle responsible for establishing the hierarchy is already there in form of the massless graviton. \\

\section*{Acknowledgements}
It is a pleasure to thank Goran Senjanovic for discussions. 
This work was supported in part by the Humboldt Foundation under Humboldt Professorship Award, by the Deutsche Forschungsgemeinschaft (DFG, German Research Foundation) under Germany's Excellence Strategy - EXC-2111 - 390814868,
and Germany's Excellence Strategy  under Excellence Cluster Origins.

\end{document}